\documentclass[prl,amsmath,amssymb,twocolumn]{revtex4}

\usepackage{graphicx}
\usepackage{subfigure}
\usepackage{adjustbox}
\usepackage{bm}
\usepackage{color}
\usepackage{braket}
\usepackage{standalone}
\usepackage{multirow}
\usepackage{tikz}
\usepackage{mathrsfs}
\usepackage{dsfont}
\usepackage[colorlinks,bookmarks=true,citecolor=blue,linkcolor=red,urlcolor=blue]{hyperref}

\newcommand{\bea}{\begin{eqnarray}}
\newcommand{\eea}{\end{eqnarray}}

\begin{document}
\title{Symmetry-protected nodal phases in non-Hermitian systems}

\author{Jan Carl Budich$^{1}$, Johan Carlstr\"om$^{2}$, Flore K. Kunst$^{2}$, and Emil J. Bergholtz$^{2}$}

\affiliation{$^1$Institute of Theoretical Physics, Technische Universit\"{a}t Dresden, 01062 Dresden, Germany\\
$^2$Department of Physics, Stockholm University, AlbaNova University Center, 106 91 Stockholm, Sweden}
\date{\today}

\begin{abstract}
Non-Hermitian (NH) Hamiltonians have become an important asset for the effective description of various physical systems that are subject to dissipation. Motivated by recent experimental progress on realizing the NH counterparts of gapless phases such as Weyl semimetals, here we investigate how NH symmetries affect the occurrence of exceptional points (EPs), that generalize the notion of nodal points in the spectrum beyond the Hermitian realm. Remarkably, we find that the dimension of the manifold of EPs is generically increased by one as compared to the case without symmetry. This leads to nodal surfaces formed by EPs that are stable as long as a protecting symmetry is preserved, and that are connected by open Fermi volumes. We illustrate our findings with analytically solvable two-band lattice models in one and two spatial dimensions, and show how they are readily generalized to generic NH crystalline systems.
\end{abstract}

\maketitle

{\emph{Introduction.---}} Nodal materials such as Dirac semimetals and Weyl semimetals are in the spotlight of current research due to their fascinating transport properties, including ramifications of quantum anomalies \cite{weylreview,diracmaterials}. The defining property of such intriguing phases of quantum matter is the topological stability of the nodal points in their band structure which may, as in the case of a spin-degenerate Dirac semimetal, rely on the presence of certain physical symmetries---a ubiquitous scenario in the context of topological phases known as symmetry protection \cite{hasankane, qizhang, bansillindas,schnyderreview}.

Very recently, intense theoretical \cite{reviewTorres,gong,kunstedvardssonetc,koziifu,yoshidapeterskawakmi,xiong,yaosongwang,yaowang,NHchern, EPrings,carlstroembergholtz,nodal4,nodal5,leethomale,lee,leykambliokhhuangchongnori,molina,yinjiangliluchen,zhuluchen,lieu,schomerus,alvarezvargastorres,shenzhenfu,wangzhangsong,rudnerlevitov,yuce,malzard,harterleejoglekar,esaki,lieu2, kawabatahisgashikawagongashidaueda} and experimental \cite{NHarc,EPringExp,NHtransition,NHexp,NHexp2,NHlaser,asymhop1,asymhop2} efforts have been made to extend the concept of topological band structures to {\emph{non-Hermitian}} (NH) systems which play an important role in the effective description of dissipation effects in various physical situations, ranging from quasi-particles with a finite lifetime in strongly correlated solids to arrays of optical resonators subject to gain and loss \cite{reviewTorres,ganainymakriskhajavikhanmusslimanirotterchristodoulides,lujoannopoulossoljacic}. There, the notion of nodal points is generalized to {\emph{exceptional points}} (EPs) \cite{Heiss,benderboettcher} which, in sharp contrast to band touching points in Hermitian systems, reflect a defective, i.e. non-diagonalizable, non-Hermitian Hamiltonian \cite{bender,brody,rotter}. Interestingly, in NH systems the occurrence of nodal band structures qualitatively changes as compared to their Hermitian counterparts \cite{koziifu,EPrings,carlstroembergholtz,NHarc}. For example, the NH analog of a Weyl semimetal with stable EPs at isolated momenta is found in two spatial dimensions rather than three \cite{BerryDeg,koziifu,NHarc,carlstroembergholtz}. Motivated by these insights and their recent experimental verification \cite{NHarc,EPringExp}, the purpose of this work is to shed light on the role of symmetry protection of EPs in NH systems, resulting in the discovery of new symmetry protected NH phases that have no immediate Hermitian analog.

\begin{figure}[t]
\centering
\includegraphics[width=\columnwidth]{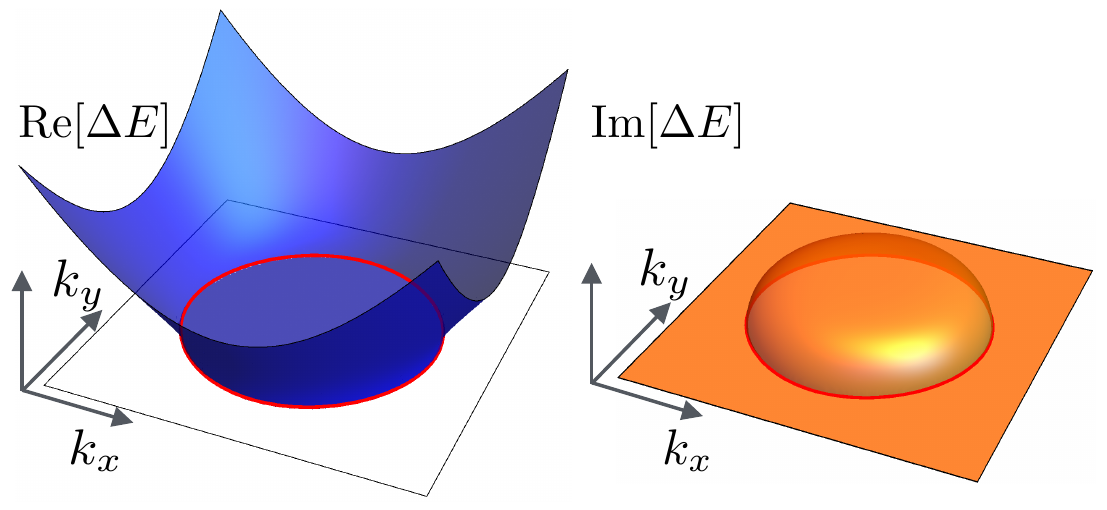}
\caption{\label{fig:one}Illustration of a generic symmetry-protected NH nodal phase, showing the real (blue) and imaginary (orange) part of the spectral gap $\Delta E$ as a function of momentum in a two-dimensional system. Contours of exceptional points are denoted by red lines, separating regions where the real part of $\Delta E$ is pinned to zero (open Fermi volumes) from those where the imaginary part of $\Delta E$ vanishes. Regions where both components are non-zero are forbidden by symmetry (see Eq.~(\ref{eq:nhsymqp})). This figure was obtained from a model of the form $H=(2-\cos k_x-\cos k_y)\sigma_x+i\sigma_z/4$ which satisfies the $Q_+$-symmetry defined in Eq.~(\ref{eq:nhsymqp}).}
\end{figure}

Below, investigating the role of genuinely NH symmetries in nodal phases, we demonstrate that symmetry protected EPs in NH band structures generically form a surface, the dimension of which is increased by one as compared to the case without symmetries, and which is qualitatively distinct from nodal surfaces known in the Hermitian context \cite{moroz}. By this mechanism, nodal NH semimetals are promoted to symmetry protected NH metals, where the surfaces of EPs form the boundaries of {\emph{open Fermi volumes}} characterized by a vanishing real part of the energy gap (see Fig.~\ref{fig:one} for an illustration). We predict that similarly to the recent observation of open Fermi arcs \cite{NHarc}, such open Fermi volumes and exceptional surfaces are directly experimentally accessible in NH photonic systems. Analyzing minimal microscopic models with two bands, we analytically prove and clarify this general principle, and, moreover, show how breaking the protecting symmetry allows one to continuously remove the EPs. 

{\emph{Symmetries  in non-Hermitian band structures.---}} The general problem of identifying the generic symmetries for NH systems, thus generalizing the celebrated ten-fold Altland-Zirnbauer classification from the Hermitian context \cite{altlandzirnbauer,schnyderryufurusakiludwig}, has been addressed by Bernard and LeClair (BL) \cite{bernardleclair, bernardleclair2}. Instead of ten symmetry classes, BL found a system of 43 classes, an extension based on additional symmetries that would not be generic in the Hermitian realm. These classes have been discussed in the context of edge states by Esaki {\it et. al.} \cite{esaki} and steps toward classifying NH phases in terms of the BL classes were very recently taken by Lieu \cite{lieu2}. Here, we start by focusing on a concrete example of such a genuinely NH symmetry coined $Q_+$ to reveal and introduce novel symmetry protected NH {\it nodal phases}. The action of $Q_+$ on the NH Hamiltonian of a system is defined as

\begin{align}
\quad H & = q H^\dagger q^{-1}, \quad q^\dagger q^{-1} = q q^\dagger= \mathbb{I}. \label{eq:nhsymqp}
\end{align}
Clearly, in the limit of a Hermitian Hamiltonian, $Q_+$ reduces to an ordinary unitary symmetry that commutes with the Hamiltonian, and hence is not considered in the Altland-Zirnbauer classification. 

{\emph{Symmetry protected EPs in two-band models.---}}
We now illustrate the idea of symmetry protected EPs on the basis of lattice periodic NH two-band models, noting that, as we demonstrate below, our results can readily be generalized beyond this minimal framework. In reciprocal space, the Bloch Hamiltonian of a NH two-band model at lattice momentum $k$ reads as
\begin{equation}
H(k) = {\bf d}(k) \cdot \boldsymbol\sigma + d_0(k) \sigma_0, \label{eq:generictwobandham}
\end{equation}
with the standard Pauli matrices $\boldsymbol\sigma$, the  $(2 \times 2)$ identity matrix $\sigma_0$, and ${\bf d} \equiv {\bf d}_R + i {\bf d}_I$ with ${\bf d}_R, {\bf d}_I \in \mathbb{R}^3$ parameterizing the Hermitian and anti-Hermitian parts of the Hamiltonian, respectively. Here and in the following, we drop the momentum dependence of ${\bf d}$ for notational brevity. The eigenvalues of the Bloch Hamiltonian are given by $E_\pm = d_0 \pm \sqrt{d_R^2 - d_I^2 + 2 i {\bf d}_R \cdot {\bf d}_I}$. Hence, EPs appear when
\begin{equation}
d_R^2 - d_I^2 = 0, \qquad {\bf d}_R \cdot {\bf d}_I = 0
\label{eq:EPcondition}
\end{equation}
are simultaneously satisfied. To find (second-order) EPs, we thus need to tune \emph{two} parameters instead of the usual three parameters in the case of Hermitian systems. This is the basic reason why nodal NH phases, even in the absence of symmetries, occur in one dimension lower than their Hermitian counterparts.

Taking into consideration the symmetry relation $Q_+$ in Eq.~(\ref{eq:nhsymqp}), we find two inequivalent possibilities for $q$, either $q = \sigma_0$ or $q$ equals one of the Pauli matrices. In the first case, Eq.~(\ref{eq:nhsymqp}) reduces to $H = H^\dagger$,  constraining our Hamiltonian to the Hermitian realm, which is not the subject of our present interest. In the second case, we take $q = \sigma_x$, such that Eq.~(\ref{eq:nhsymqp}) takes the explicit form $H = \sigma_x H^\dagger \sigma_x$. This leads to the following constraints for the Hamiltonian in Eq.~(\ref{eq:generictwobandham})
\begin{equation}
d_x, d_0 \in \mathbb{R}, \qquad d_y, d_z \in i \mathbb{R}.
\label{eq:symconcrete}
\end{equation}
This means that the relation ${\bf d}_R \cdot {\bf d}_I = 0$ in Eq.~(\ref{eq:EPcondition}) is \emph{trivially} satisfied. Therefore, to find EPs in this case, we only need to solve $d_R^2 - d_I^2 = 0$, and we thus need to tune only \emph{one} parameter instead of two. This observation has two important consequences. First, EPs, exceptional lines and exceptional surfaces generically appear in one-, two- and three-dimensional systems, respectively. This amounts to a \emph{further reduction} by one of the spatial dimension in which nodal phases occur as compared to the NH case without symmetries. Second, the eigenvalue equation in the presence of $Q_+$ reduces to
\begin{equation}
E_\pm = \pm \sqrt{d_R^2 - d_I^2},
\label{eq:epm}
\end{equation}
where we have neglected $d_0$ as this term does not affect the gap $\Delta E$ between the bands. 

The implication of Eq.~(\ref{eq:epm}) is that the parameter space is divided into regions where the energy is either real, or purely imaginary, and the exceptional points thus form boundaries between Fermi volumes of vanishing energy, and  regions of infinite quasi particle lifetimes where ${\rm{Im}}(E)=0$ 
(see Fig.~\ref{fig:one} for an illustration). The inclusion of $d_0$ would just move these Fermi volumes away from ${\rm Re}(E) = 0$, but the energy gap $\Delta E$ would still have a vanishing real part.

We now elaborate on these insights with the help of concrete microscopic lattice models in one spatial dimension (1D) and in 2D on a square lattice with unit lattice constant. To this end, we consider
\begin{align}
{\bf d} = \left(m+1-\cos(k_x)-\cos(k_y),i\sin(k_x),i\sin(k_y)\right),
\label{eq:toy2D}
\end{align}
where $m\in \mathbb R$. Clearly the Hamiltonian $H=\bf d \cdot \boldsymbol{\sigma}$ resulting from Eq.~(\ref{eq:toy2D}) satisfies the symmetry relation $Q_+$ with $q=\sigma_x$ (see Eq.~(\ref{eq:nhsymqp}) and Eq.~(\ref{eq:symconcrete})). To obtain our 1D model, we simply set $k_y=0$ for now. If $m=0$ as well, EPs occur whenever $\cos(k_x)=\pm \sin(k_x)$, i.e. at $k_x=\pi/4 (\text{mod} \pi/2)$. For $m \ne 0$, EPs satisfying the condition $\left(m-\cos(k)\right)^2-\sin(k)^2=0$ (see Eq.~\ref{eq:EPcondition}) are still found for $\lvert m\rvert \le \sqrt{2}$, and the spectrum becomes fully gapped only for $\lvert m\rvert>2$. Furthermore, it is clear that EPs at which the spectrum of $H$ switches from being purely real to purely imaginary are always connected via the aforementioned open Fermi volumes, which in the present 1D case form open arcs. Thus, as long as the symmetry $Q_+$ is preserved, the EPs are stable in the sense that they can only be removed pairwise by bringing them together in momentum space and contracting the arcs. This behavior exemplifies and corroborates our claim of the existence of a stable nodal NH phase protected by the symmetry $Q_+$. To explicitly demonstrate the importance of this symmetry protection, we add a {\emph{symmetry breaking}} perturbation $H_b= i\delta \sigma_x$ with $\delta>0$. Then, the overlap ${\bf d}_R \cdot {\bf d}_I$ becomes finite and EPs can be continuously removed already at arbitrarily small $\delta$ (cf. Eq.~(\ref{eq:EPcondition})).

\begin{figure*}[t]
\includegraphics[width=\linewidth]{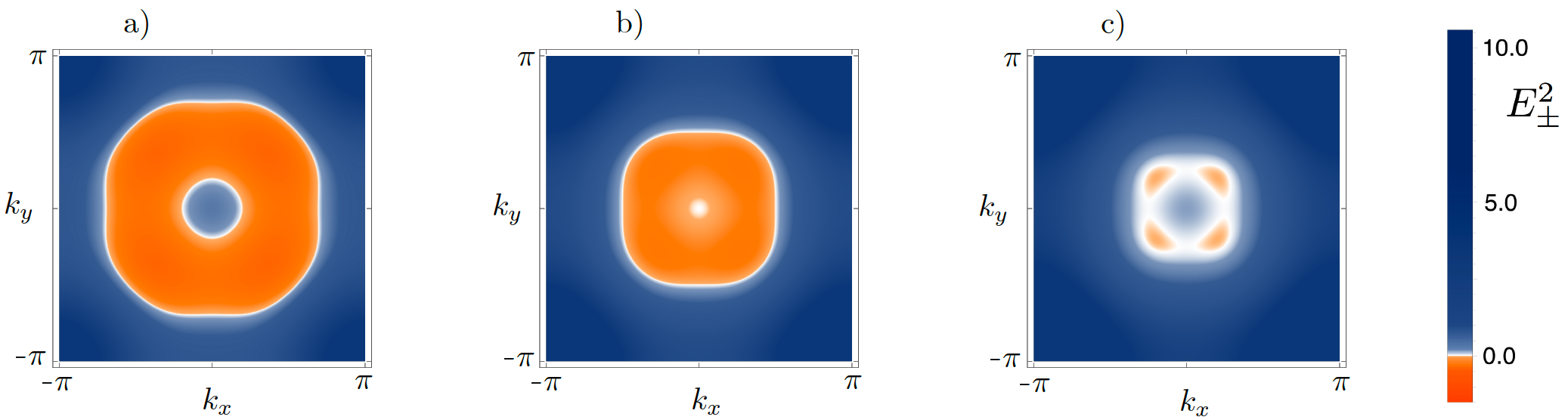}
\caption{Momentum dependence of the squared energy $E_\pm^2$ spectrum of the model defined in Eq.~(\ref{eq:toy2D}), which is real due to Eq.~(\ref{eq:epm}) and thus contains the full information about the energies $E_\pm$. The white contours hallmark the lines of EPs, separating regions with real energies $E_\pm$ (blue, positive $E_\pm^2$) from the open Fermi volumes characterized by $\textrm{Re}(E_\pm)=0$ (orange, negative $E_\pm^2$). The plot parameters from left to right are $m=0.25,~m=1.0,~m=1.42$. At $m=1.0$ (panel b), the inner line of EPs collapses to a single Hermitian nodal point. At $m=1.42$ (panel c), the EPs form a pattern reminiscent of Fermi pockets with a fourfold rotational symmetry.}
\label{fig:2deps}
\end{figure*}

Proceeding to the 2D case, we now also allow $k_y$ to take arbitrary values in the first Brillouin zone, i.e. $k_y \in [-\pi,\pi]$ in Eq.~(\ref{eq:toy2D}). A good quantity to understand the spectrum of our model is the squared energy $E_\pm^2$, which is real due to ${\bf d}_R \cdot {\bf d}_I = 0$, and has zeros at the EPs, where it switches from being positive (real $E_\pm$) to negative (purely imaginary $E_\pm$). In Fig.~\ref{fig:2deps}, we show the momentum dependence of $E_\pm^2$, for various values of $m$ demonstrating the rich physical phenomenology of our model (\ref{eq:toy2D}): For $\lvert m+1 \rvert < 1+\sqrt{2}$,  the spectrum exhibits a pair of closed lines of EPs connected by open 2D Fermi surfaces (see Fig.~\ref{fig:2deps}a), interrupted by the point $\lvert m+1\rvert =2$, where one line of EPs collapses to a single Hermitian nodal point (see Fig.~\ref{fig:2deps}b). When further increasing $\lvert m+1\rvert$ beyond $1+\sqrt{2}$, these lines of EPs split into four NH Fermi pockets  (see Fig.~\ref{fig:2deps}c) and eventually disappear at $\lvert m+1\rvert= \sqrt{6}$, where the spectrum becomes fully gapped for all $\lvert m+1\rvert>\sqrt{6}$. Again, if we add the perturbation $H_b$, all EPs are removed at arbitrarily small $\delta$, highlighting the fact that the stability of the EP's hinges on symmetry.

It is worth emphasizing that all the ingredients considered here, in particular all terms in our model (\ref{eq:toy2D}), are available with state of the art experimental techniques: An imaginary $d_z$ is relatively easily achieved by staggered gain and loss terms acting on the two sublattices of the system \cite{NHexp,NHexp2}, and asymmetric, thus anti-Hermitian, hopping terms corresponding to imaginary $d_y$ (and $d_x$) have recently been realized in experiments \cite{asymhop1,asymhop2}.

%%%%%%%%%%%%%%%%%
%% CASE OF MANY BANDS %%
%%%%%%%%%%%%%%%%%

\begin{figure}[!htb]
 \hbox to \linewidth{ \hss
\includegraphics[width=\linewidth]{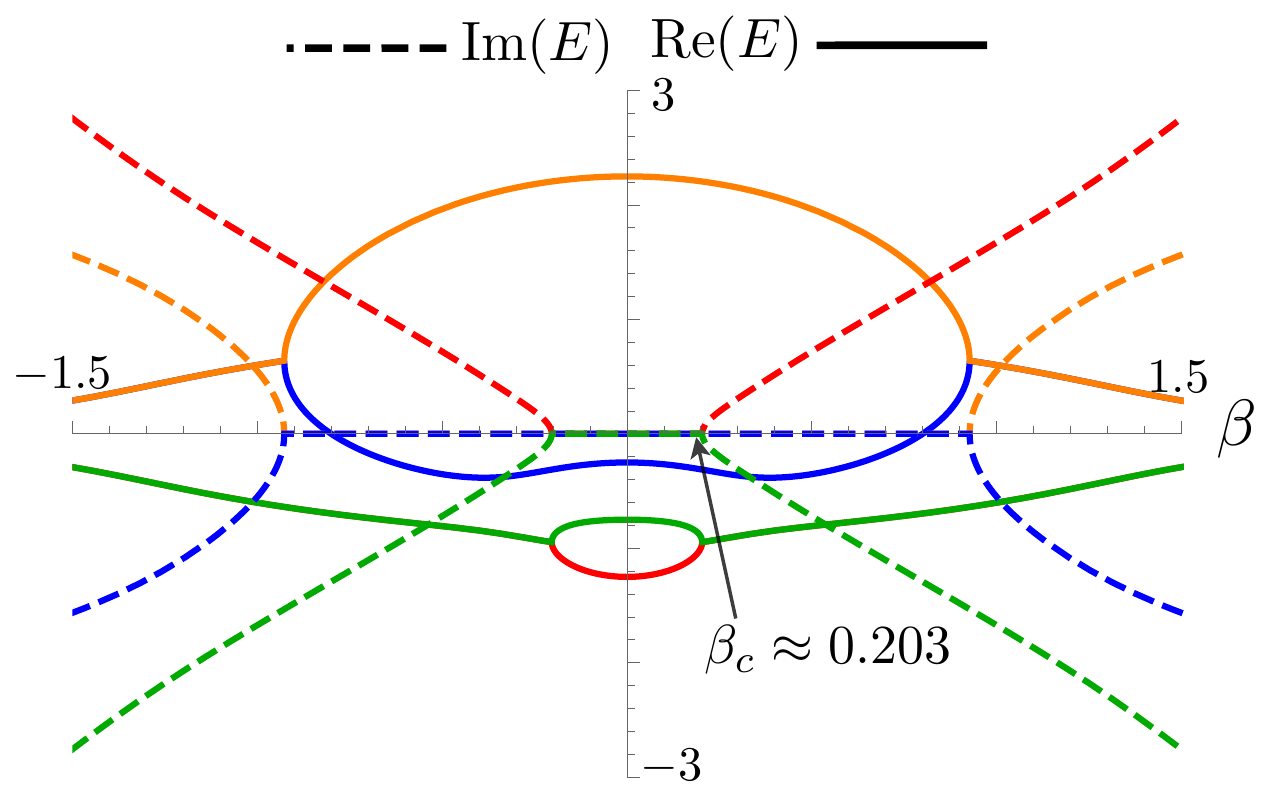}
 \hss}
\caption{
Real (solid) and imaginary (dashed) parts of the spectrum of a generic four band model subject to the symmetry constraint  Eq.~(\ref{Q-sym}), with $q=\tau_z\sigma_0$ and $\epsilon=+1$ and thus the spectral symmetry (\ref{Isymmetry}). Each color corresponds to a particular band. The Hamiltonian is $H=\alpha H_A+i\beta H_B$ as in  Eq.~(\ref{Hdecomp}) with the choice $H_A=3/4\;\tau_z\sigma_z+1/2\;\tau_z\sigma_0+\;\tau_0\sigma_z$ and $H_B=\;\tau_x\sigma_x+\;\tau_y\sigma_x+\;\tau_y\sigma_0$. The parameter $\alpha=1$ is taken to be constant while $\beta$ is varied on the horizontal axis. In the limit  $\beta\to 0$, the model becomes Hermitian and for $|\beta|\le\beta_c\approx 0.203$, the spectrum is entirely real and non-degenerate, whilst $\beta\to \pm\infty$ corresponds to the anti-Hermitian limit with imaginary eigenvalues. Between these limiting cases, pairs of energy levels coalesce at exceptional points separating regions with entirely real and imaginary gaps as dictated by Eq.~(\ref{Isymmetry}).
}
\label{spectrum}
\end{figure}

{\emph{Generalization to many bands and other symmetries.---}} 
We now discuss the generalization of our results beyond the minimal setting of two band models as well as to other symmetries. Building upon the BL classification \cite{bernardleclair, bernardleclair2}, we consider the following system of generic NH symmetries:
\begin{align}
P: \quad H &= -pHp^{-1}, \quad p^2 = \mathbb{I}, \label{P-sym}\\
C: \quad H & = \epsilon_c c H^T c^{-1}, \quad c^T c^{-1} = \pm \mathbb{I}, \label{C-sym}\\
K: \quad H&= \epsilon_k k H^* k^{-1}, \quad k k^* = \pm \mathbb{I}, \label{K-sym} \\
Q: \quad H & = \epsilon_q q H^\dagger q^{-1}, \quad q^\dagger q^{-1} = \mathbb{I}, \label{Q-sym}
\end{align}
where $p,\;c,\;k,\;q$ are understood to be unitary transformations, and $\epsilon_i=\pm 1,~i=c,k,q$. To derive the implications of  Eq.~(\ref{P-sym}-\ref{Q-sym}) for the EPs of a NH system, it is helpful to consider the corresponding spectral symmetries. Specifically, we have
$|H- E|=0\implies |\epsilon_i \Gamma H' \Gamma^{-1}-E |=0 \implies | \epsilon_i H'-E |=0\implies |  H-\epsilon_i E' |=0$, where we have used the fact that $\Gamma=p,\;c,\;k$ or $ q$ is unitary, and where $H'$, depending on which symmetry is considered, denotes  $H,\;H^T,\; H^* $, and $H^\dagger$, respectively. This in turn gives rise to three principal spectral symmetries:
\begin{align}
 \{E_i\} \; &=&\; \{-E_i\} & : & P \text{ and } C,\;\epsilon_c=-1 		\label{Fullsymmetry}\\
 \{E_i\} \; &=&\; \{E_i^*\} & : & K,\;\epsilon_k=1 \text{ and } Q,\;\epsilon_q=1 		\label{Isymmetry}\\
 \{E_i\} \; &=&\; \{-E_i^*\} & : & K,\;\epsilon_k=-1 \text{ and } Q,\;\epsilon_q=-1 		\label{Rsymmetry}
\end{align}
where $\{E_i\}$ denotes the set of (complex) energy eigenvalues. The constraint in Eq.~(\ref{Fullsymmetry}) merely implies that the eigenvalues are symmetric around zero. This is in general not sufficient for the occurrence of symmetry protected EPs, as is immediately clear from the two-band case, where it simply forbids terms that break particle-hole symmetry, i.e. the presence of $d_0(k)$ in Eq.~(\ref{eq:generictwobandham}). Similarly, as noted in the preceding two-band example, there exist (trivial) representations of the unitaries $q$ and $k$ such that Eq.~(\ref{K-sym}) and Eq.~(\ref{Q-sym}) simply force the model to be purely Hermitian or anti-Hermitian, respectively, e.g. for unitaries that simply (anti-)commute with the Hamiltonian. Here, however, we focus on genuinely non-Hermitian cases leading to non-trivial constraints as well as exceptional points. We emphasize that the case of Eq.~(\ref{Isymmetry}), where the spectrum is invariant under complex conjugation, is also relevant for NH systems with $\mathcal{P}\mathcal{T}$ symmetry \cite{Heiss}. This gives rise to generic exceptional sets of dimensionality $D-1$ as we show in the following:
We can decompose the NH Hamiltonian into a Hermitian and an anti-Hermitian part according to
\bea
H= \alpha H_A +i \beta H_B,\;H_A=H_A^\dagger,\; H_B=H_B^\dagger \label{Hdecomp}
\eea
 where $\beta = 0$ corresponds to the Hermitian limit where all eigenvalues must be real. 
 Next we note, that if the spectrum of $H_A$ is non-degenerate, i.e.
$|E_i-E_j|\ge \Delta_A>0,\;i\not=j$,
then the eigenvalues of $H$ must remain real in the proximity of $\beta=0$. This is because the spectrum is a continuous function of model parameters, and complex conjugate pairs of eigenvalues cannot continuously emerge from a set of non-degenerate real energies with gaps bounded by $\Delta_A$. Hence, there exists a finite $\beta_c>0$ such that all eigenvalues are real for $|\beta|\le \beta_c$, see  Fig.~\ref{spectrum} for an illustration.

If we on the other hand consider the anti-Hermitian limit $\alpha\to 0$, while $\beta$ remains finite, then the spectrum becomes purely imaginary. By analogy to the Hermitian limit, if
the the spectrum is non-degenerate, i.e. $|E_i-E_j|\ge \Delta_I>0$, then these gaps must remain open for a finite range
 $|\alpha|\le \alpha_c$.

It thus follows that there exist finite neighborhoods in parameter space where pairs of eigenvalues are either real, or complex and related by conjugation respectively. At the boundaries between these regions, both the real and imaginary parts of a gap between two of the energy levels must necessarily vanish, meaning that second order exceptional points form $D-1$ dimensional surfaces that are topologically stable in the presence of the spectral symmetry (\ref{Isymmetry}). This generalizes the above discussion of two-band models to generic NH band structures with a symmetry of the form (\ref{Isymmetry}).

Finally, we note that the spectral symmetry (\ref{Rsymmetry}), where the set of eigenvalues  is odd under complex conjugation, has similar consequences on the occurrence of symmetry protected EPs as that of (\ref{Isymmetry}). This is because for a Hamiltonian $H$ satisfying Eq.~(\ref{Isymmetry}), it is clear that $iH$ satisfies Eq.~(\ref{Rsymmetry}). Hence, the EPs once again form $D-1$ dimensional surfaces that separate domains in parameter space. However, in the case of Eq.~(\ref{Rsymmetry}), the EPs are either characterized by purely imaginary eigenvalues, or complex eigenvalues that appear in pairs such that ${\rm Re}(E_i)=-{\rm Re}(E_j)$.

{\emph{Concluding discussion.--- }} In this work, we have investigated the impact of symmetries on nodal non-Hermitian band structures. Remarkably, even symmetries that are redundant in the Hermitian limit are shown to have a profound impact on NH band structures; they reduce the number of parameters that need to be tuned in order to find generic EPs from two to one, hence separating them even further from the generic nodal points in Hermitian systems that required three tuning parameters. The $Q_+$ symmetry in Eq.~(\ref{eq:nhsymqp}) provides an emblematic example of this where the Hermitian limit is trivial while the non-Hermitian theory is very rich with generic symmetry protected EPs as we discussed in one and two spatial dimensions. In particular, the EPs are generally accompanied by what we dub Fermi volumes: open regions of vanishing real part of the energy gap which have the same dimension as the system itself. These Fermi volumes may be readily observed with scattering experiments similar to those that recently reported on the observation of two-dimensional bulk Fermi arcs in photonic crystals \cite{NHarc}. 
Our systematic approach does not only reveal striking features of band topology, but also puts well established knowledge into a coherent theoretical framework, for instance the ubiquitous occurrence of EPs in $\mathcal{P}\mathcal{T}$ symmetric systems where tuning of only a single parameter is enough to generically probe an exceptional point. 
  
The enriching effects of non-Hermiticity found here for nodal systems stand in strong contrast to the outcome of a very recent study in the context of gapped NH band-structures \cite{gong}, where it has been found that the standard Altland-Zirnbauer classification \cite{altlandzirnbauer} is reduced from ten to seven distinct classes, and an accompanying sharp drop in the number of gapped topological phases, including their complete absence in two dimensions has been reported. That we instead find non-Hermiticity to make the problem richer stems from the fact that we consider inherently non-Hermitian symmetries, namely the Bernard-LeClair classes, and that we study nodal rather than gapped phases.

Finally, we note that the ingredients needed to create and manipulate the nodal phases discussed here are already available in a number of experimental settings ranging from photonic crystals, to micro mechanical resonators and arrays of classical waveguides. Hence, our work provides a basis for the experimental search of such phases as well as for further in-depth theoretical explorations of the rich variety of nodal NH phases. 

{Note added:} After a preprint of our manuscript was made available online, several other preprints appeared that elaborate on various aspects of symmetry protected nodal NH phases \cite{Okugawa,Yoshida,Zhou}.

\acknowledgments
{\it Acknowledgments.---} 
F.K.K., J.C., and E.J.B. are supported by the Swedish Research Council (VR) and the Wallenberg Academy Fellows program of the Knut and Alice Wallenberg Foundation. J.C.B. acknowledges financial support from the German Research Foundation (DFG) through the Collaborative Research Centre SFB 1143.

\end{document}